\RequirePackage[l2tabu, orthodox]{nag}
\RequirePackage{snapshot}

\documentclass[9pt,onecolumn]{extarticle}

\makeatletter
\if@twocolumn
  \usepackage[dvips,letterpaper,top=0.5in, bottom=0.5in, left=0.75in, right=0.5in,includefoot,heightrounded]{geometry}
\else
  \usepackage[dvips,letterpaper,margin=1in,includefoot,heightrounded]{geometry}
\fi

\usepackage{srcltx}

\usepackage[russian,portuges,english]{babel}

\iflanguage{portuges}
    {\newcommand{\keywordname}{Palavras-chaves}}
    {\newcommand{\keywordname}{Keywords}}

\usepackage{amsmath}
\usepackage{amssymb,amsfonts}

\usepackage{abstract}

\usepackage{graphicx}
\usepackage[usenames,dvipsnames,svgnames,x11names]{xcolor}
\usepackage{subfigure}
\usepackage{tikz}

\usepackage{booktabs}

\usepackage{setspace}
\usepackage{flushend}

\usepackage{cite}

\usepackage{hyperref}
\usepackage[normalem]{ulem}

\usepackage{enumerate}

\usepackage[short,12hr]{datetime}
       \usepackage{fouriernc}
\makeatletter

\newcommand{\printtitle}{%
\makeatletter
\if@twocolumn

\twocolumn[%
  \maketitle
  \begin{onecolabstract}
    \myabstract
  \end{onecolabstract}
  \begin{center}
    \small
    \textbf{\keywordname}
    \\\medskip
    \mykeywords
  \end{center}
  \bigskip
]
\saythanks
\else
  \maketitle
  \begin{onecolabstract}
    \myabstract
  \end{onecolabstract}
  \begin{center}
    \small
    \textbf{\keywordname}
    \\\medskip
    \mykeywords
  \end{center}
  \bigskip
  \onehalfspacing
\fi
\makeatother
}

\author{%
H. M. de Oliveira
\thanks{Departamento de Estat\'{\i}stica, Universidade Federal de Pernambuco, Recife, PE,
E-mail: \url{hmo@de.ufpe.br}}
\and
R.~J.~Cintra%
\thanks{%
Departamento de Estat\'{\i}stica, Universidade Federal de Pernambuco, Recife, PE,
E-mail: \url{rjdsc@de.ufpe.br}}
}

\title{%
Geometrical Representation for Number-theoretic Transforms}

\newcommand{\myabstract}{%
This short note introduces a geometric representation
for binary (or ternary) sequences.
The proposed representation
is linked to
multivariate data plotting
according to
the radar chart.
As an illustrative example, the binary Hamming transform recently proposed is geometrically interpreted.
It is shown that codewords of standard Hamming code $\mathcal{H}(N=7,k=4,d=3)$ are invariant vectors under the Hamming transform.
These invariant are eigenvectors of the binary Hamming transform.
The images are always inscribed in a regular polygon of unity side, resembling triangular rose petals and/or ``thorns''.
A geometric representation of the ternary Golay transform, based on the extended Golay $\mathcal{G}(N=12, k=6, d=6)$ code over $\operatorname{GF}(3)$ is also showed.
This approach is offered as an alternative representation of finite-length sequences over finite prime fields.
}

\newcommand{\mykeywords}{%
Finite fields, Hamming binary transforms, Golay ternary transforms, geometric representations.
}

\date{\today\ @ \currenttime}

\date{}

\begin{document}

\printtitle

\section{Introduction}

Discrete transforms defined over a finite field
are signal processing tools
capable of providing
Fourier analysis~\cite{pollard1971fast,harvey2014faster,mert2020extensive}
while
operating
in error-free structure.
Because its arithmetic is performed in a finite field,
fixed-point implementations can provide exact computation
and
simple hardware requirements.
Several signal processing contexts
were benefited by finite fields transforms \cite{cooklev1994theory,jullien1987complex,
hegde2020medical,
toivonen2005video,
ozcan2019high}
with
applications linked to
the computation of the discrete convolution
by means of modular arithmetic
and
to image processing methods~\cite{boussakta1999number,gertner1990image}.
Number-theoretic transforms (NTTs) are finite-field transforms
that operate over $\operatorname{GF}(p)$, where $p$ is a prime number,
as opposed to operating over the extension field
$\operatorname{GF}(q)$, where $q$ is a power of a prime.
Such particular results in simple, error-free architectures
while preserving an analogy to real-valued computation.

Besides their applications in signal processing,
NTTs have been linked to error correcting codes.
Based on the Fourier NTT and the Hartley NTT,
the Fourier and Hartley codes were
introduced~\cite{campellodesouza2011codigos,oliveira2007fourier}.
Conversely,
popular error-correcting codes~\cite{lin2004error},
such as
the Hamming~\cite{hamming1950error}
and
Golay codes~\cite{Macwilliams1977theory},
inspired the
introduction of
the Hamming number-theoretic transform (HamNTT)~\cite{paschoal2018hamming}
and
the Golay number-theoretic transform (GNTT)~\cite{paschoal2018hamming}
which
extend the theory introduced in~\cite{souza2009fourier,paschoal2018novas}.
In fact,
an isomorphism between linear codes and transforms
was identified in~\cite{paschoal2018hamming}.

The goal of this paper is
to introduce a representation for
sequences over $\operatorname{GF}(p)$
as
a tool for the investigation of
number-theoretic transforms.

\section{Geometric representation}

Let
$\operatorname{GF}(p)$
be a Galois field of order $p$,
where $p$ is a prime number.
A message of length $N$
is a sequence
$\mathbf{x}=[x_0, x_1, \ldots,x_{N-1}]$
such
that
$x_i \in \operatorname{GF}(p)$,
$i=0,1,\ldots,N-1$.
Based on the radar representation~\cite{albo2016radar}
(also referred to as web chart or spider chart),
we
propose
a geometrical representation
for
such messages.
The geometric representation
consists
of mapping the message symbols
in points on the complex plane
according to
the
following expression:
\begin{align}
\label{equation-vertices}
z_k
=
x_k \pmod{p}
\cdot
\exp\left(j\frac{2\pi}{N}k \right)
,
\quad
k = 0, 1, \ldots, N-1
.
\end{align}
The set of points
$\{z_0, z_1, \ldots, z_{N-1} \}$
defines
a constellation
on which
a geometric shape
composed of polygons and segments
can be derived.

\begin{figure}
\centering
\subfigure[]{\input{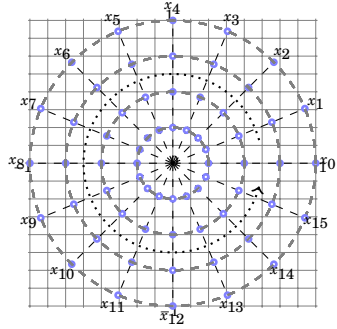}}
\caption{Constellation for $N=16$ and $p=5$.
Each radial axis corresponds to a message symbol.
The arrow indicates the symbol ordering.}
\label{figure-general-constellation}
\end{figure}

The geometric representation
is constructed according to the following procedure:

\begin{enumerate}
\item
Locate on the complex plane the loci
of the $N$th roots of the unity
$e^{j\frac{2\pi}{N}k}$,
$k = 0, 1, \ldots, N-1$;

\item
Scale each root of the unity
by the corresponding
symbol $x_k$
as shown in~\eqref{equation-vertices}
and
plot the resulting point $z_k$ on the plane;

\item
Draw line segments joining
the
points
$z_{k+1}$ and $z_k$
to obtain the geometric representation.

\end{enumerate}
For instance,
Figure~\ref{figure-general-constellation}
shows the required constellation for
16-point messages over $\operatorname{GF}(5)$.

\section{Geometrical Representation of Binary Sequences}

\begin{figure}
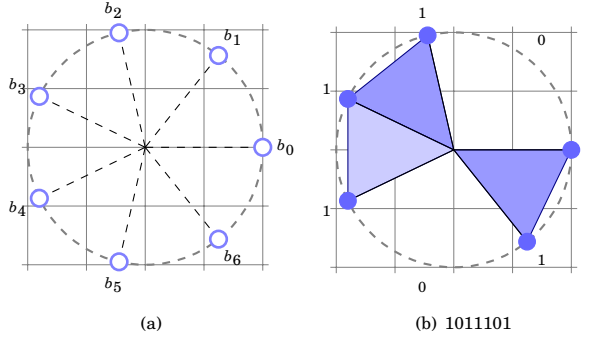

\centering
\subfigure[]{\input{binary_flower_scheme_7.tikz}}
\subfigure[1011101]{\input{1011101.tikz}}
\caption{Geometric representation of codewords.
(a) Geometric space.
Circles are filled or not according to the bits of the word.
(b) Representation of the word $[1011101]$.}
\label{fig:geom}
\end{figure}

\begin{figure}
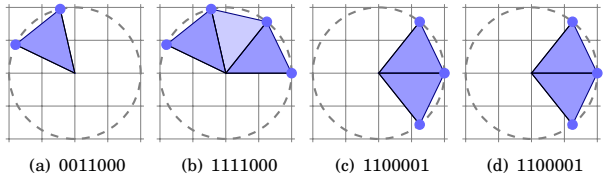

\centering
\subfigure[0011000]{\input{0011000-small.tikz}}
\subfigure[1111000]{\input{1111000-small.tikz}}
\subfigure[1100001]{\input{1100001-small.tikz}}
\subfigure[1100001]{\input{1100001-small.tikz}}

\caption{(a)-(b) A sequence and its associate Hamming NTT sequence.
(c)-(d) An invariant sequence to the Hamming NTT.}
\label{figure-hamming}
\end{figure}

The special case
$p=2$ and $N=7$
is
suitably
linked to the Hamming NTT~\cite{paschoal2018hamming}.
Based
on the binary Hamming code
$\mathcal{H}(7, 4, 3)$,
we get the $7 \times 7$ binary Hamming NTT,
whose transformation matrix is~\cite{paschoal2018novas}
\[
\mathbf{T}_\text{HamNTT}
=
\begin{bmatrix}
0&1&0&1&1&0&0\\
1&0&1&0&0&1&0\\
1&0&0&1&0&0&1\\
0&0&0&1&0&0&0\\
0&0&0&0&1&0&0\\
0&0&0&0&0&1&0\\
0&0&0&0&0&0&1\\
\end{bmatrix}
.
\]
This transform has the property that
its eigenvector matrix
is equal to
the
generator matrix of the code
$\mathcal{H}(7, 4, 3)$,
i.e.:
\[
\operatorname{eig}
\left\{
\mathbf{T}_\text{HamNTT}
\right\}
=
\begin{bmatrix}
1&1&0&0&0&0&1\\
1&1&1&0&0&1&0\\
1&0&1&0&1&0&0\\
0&1&1&1&0&0&0\\
\end{bmatrix}
=
\mathbf{G}
.
\]
Input data is represented
by
$\mathbf{x} = [b_0, b_1, \ldots,b_6]$,
where
$b_k \in \{ 0, 1 \}$,
$k=0,1,\ldots,6$.
Such sequence is used to create small circles on the dashed circunference
shown in Figure~\ref{fig:geom}(a).
If $b_k = 1$, then
a small circle filled in color is placed
at position $z_k= e^{j\frac{2\pi}{7} k}$;
otherwise,
if $b_k = 0$, then the small circles are not generated.
The next step is the petal creation:
any two consecutive filled circles
forms a triangle with the origin $(0,0)$
producing a petal
(alternately shaded in light and dark color).
Points $b_0$ and $b_6$
(cyclical geometry) are understood as neighbors.
Figure~\ref{fig:geom}(b)
represents the byte $[1011101]$.
The above linear transform maps 7-bit sequences over 128 possible patterns.
In the Appendix,
Figure~\ref{figure-binary-panel}
lists all 7-bit sequences in the proposed representation.
Figure~\ref{figure-hamming}(a)-(b) shows a particular sequence and
its associate transformed sequence according to the HammNTT.
Some sequences are invariant to the  Hamming NTT
such as
$\mathbf{x} = [1100001]^\top$
which satisfies
$\mathbf{T}_\text{HamNTT} \cdot \mathbf{x} = \mathbf{x}$.
Figure~\ref{figure-hamming}(c)-(d) displays an invariant sequence
and its transformed sequence.

\begin{figure}[!ht]
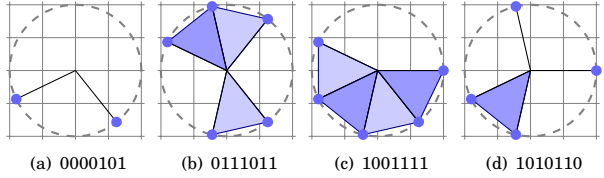

\centering

\subfigure[0000101]{\input{0000101-small.tikz}}
\subfigure[0111011]{\input{0111011-small.tikz}}
\subfigure[1001111]{\input{1001111-small.tikz}}
\subfigure[1010110]{\input{1010110-small.tikz}}

\caption{Geometric representation
(a) only thorns,
(b)-(c) petals,
and
(d) thorns and petals.
}
\label{fig:thorns}
\end{figure}

\section{Geometrical Representation of the Ternary Golay Transform}

For the ternary Golay codes, the extended Golay code has parameters $\mathcal{G}(N=12, k=6, d=6)$ over $GF(3)$.
The new geometric space can be constructed
by taking now a new ensemble of ``representative complex points''
according to:
\begin{align}
q_k
=
r_i
\cdot
\exp
\left(
j
\frac{2\pi}{12}k
\right)
,
\quad
i = 0, 1, 2;
\
k = 0, 1, \ldots 11
,
\end{align}
where
$r_i = i$.
Noticing that
$2 \equiv -1 \mod 3$,
we can write
the associate
Golay NTT matrix as follows:
\begin{equation*}
\arraycolsep=2pt
\mathbf{T}_\text{EG}^{(1)}  =
\left[\begin{array}{rrrrrrrrrrrr}
1&-1&-1&-1&-1&-1&\phantom{-}1&\phantom{-}0&\phantom{-}0&\phantom{-}0&\phantom{-}0&\phantom{-}0\\
-1&1&-1&1&1&-1&0&1&0&0&0&0\\
-1&-1&1&-1&1&1&0&0&1&0&0&0\\
-1&1&-1&1&-1&1&0&0&0&1&0&0\\
-1&1&1&-1&1&1&0&0&0&0&1&0\\
-1&-1&1&1&-1&1&0&0&0&0&0&1\\
-1&-1&1&0&0&1&-1&1&0&0&0&0\\
-1&1&-1&1&0&0&1&1&1&0&0&0\\
-1&0&1&-1&1&0&1&0&1&1&0&0\\
-1&0&0&1&-1&0&1&0&0&1&1&0\\
-1&1&0&0&1&-1&1&0&0&0&1&1\\
1&-1&-1&0&-1&0&0&1&1&0&0&1\\
\end{array}\right]
.
\end{equation*}
The above matrix can be efficiently implemented in hardware
since it does not require any multiplication operations
as its entries are over $\{0, \pm 1\}$.
Thus
the Golay NTT requires only additions in order to be computed
and
it
is applicable to any sequence of the $\{\operatorname{GF}(3)\}^{12}$-space.

\begin{figure}[t]
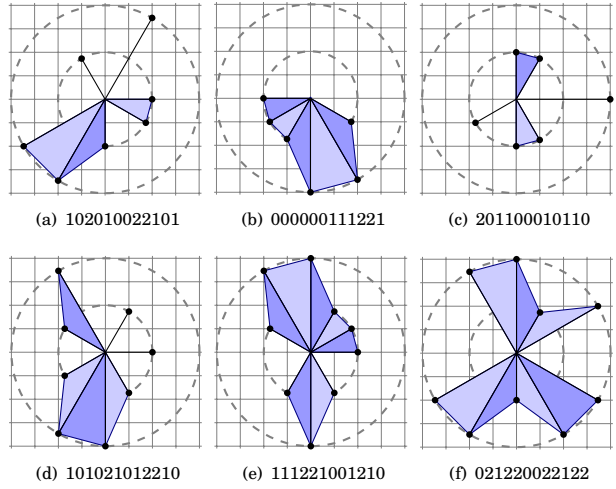

\centering

\subfigure[102010022101]{\input{./t-102010022101.tikz}}
\subfigure[000000111221]{\input{./t-000000111221.tikz}}
\subfigure[201100010110]{\input{./t-201100010110.tikz}}

\subfigure[101021012210]{\input{./t-101021012210.tikz}}
\subfigure[111221001210]{\input{./t-111221001210.tikz}}
\subfigure[021220022122]{\input{./t-021220022122.tikz}}

\caption{Golay NTT pairs.
Input data: (a), (b), and (c); transformed data: (d), (e), and (f), respectively.}
\label{fig:GolayFIG}
\end{figure}

\begin{figure}[t]
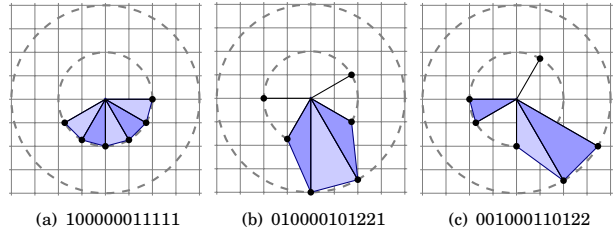

\centering

\subfigure[100000011111]{\input{./t-100000011111.tikz}}
\subfigure[010000101221]{\input{./t-010000101221.tikz}}
\subfigure[001000110122]{\input{./t-001000110122.tikz}}
\caption{Golay NTT pairs: Invariant sequences.}
\label{fig:GolayINV}
\end{figure}

Illustrative examples of the effect of
the Golay number-theoretic transform~\cite{paschoal2018hamming}
on ternary vectors of length~12 are shown in Figures~\ref{fig:GolayFIG} and \ref{fig:GolayINV}.
Note that complex symbols are always vertices of one of the two dodecagons.
Again, colors \textcolor{RoyalBlue}{light blue} and \textcolor{blue}{dark blue} are adopted alternatively in consecutive petals,
without major implications, except in improving the visualization.
Three codewords were chosen at random:
$[102010022101]$,
$[000000111221]$,
and
$[201100010110]$.
By applying the Golay NTT to these sequences,
we obtain the following
transformed sequences:
\begin{align*}
\mathbf{T}_\text{EG}^{(1)} \cdot [102010022101]^\top & =[101021012210],\\
\mathbf{T}_\text{EG}^{(1)} \cdot [000000111221]^\top & =[111221001210]^\top,\\
\mathbf{T}_\text{EG}^{(1)} \cdot [201100010110]^\top & =[021220022122]
.
\end{align*}

Invariants of the Golay NTT can the readily obtained from the generator matrix.
For instance,
the following
codewords are invariants:
$[100000011111]$,
$[010000101221]$,
and
$[001000110122]$,
as demonstrated by:
\begin{align*}
\mathbf{T}_\text{EG}^{(1)} \cdot [100000011111]^\top & =[100000011111],\\
\mathbf{T}_\text{EG}^{(1)} \cdot [010000101221]^\top & =[010000101221],\\
\mathbf{T}_\text{EG}^{(1)} \cdot [001000110122]^\top & =[001000110122]
.
\end{align*}
In the Appendix,
Figure~\ref{figure-selected-ternary-flowers}
shows
a subset of the possible words.

\section{Conclusions}

This note
introduces
a geometric representation for
finite sequences of elements defined over a finite field.
This approach provides a defiant reading for the Hamming and Golay transforms.
To the best of our knowledge,
no similar proposal to convert sequences into images was found,
which
consists
of
assigning angles to the position of the symbol in the sequence
as described in~\eqref{equation-vertices}.
Phases (angles) are meaningless, as in radar charts.
Such a representation
has
potential applications in several fields of error correcting codes and signal processing over finite fields, including:
(i)~RLE run length encoding,
(ii)~burst error correcting codes,
(iii)~binary SP,
and
(iv)~theory of filter banks.
The proposed approach can lead to new insights and interpretations
in the design of
coding and signal processing methods
dedicated to sequences over finite fields.
As future research,
we
aim at deriving
extended versions of the Hamming or Golay codes,
which are \emph{self-dual} codes \cite{conway2013sphere}.
Such codes could be employed
to obtain
new number-theoretic transforms.
As shown in the Appendix,
several geometric and symmetry patterns arise
that can be further investigated.
Such symmetries might lead to a
better understanding of practical issues
in programming
and
in hardware implementation
linked to the discussed codes.

\appendix

\section{Geometric Representations}

\begin{figure*}
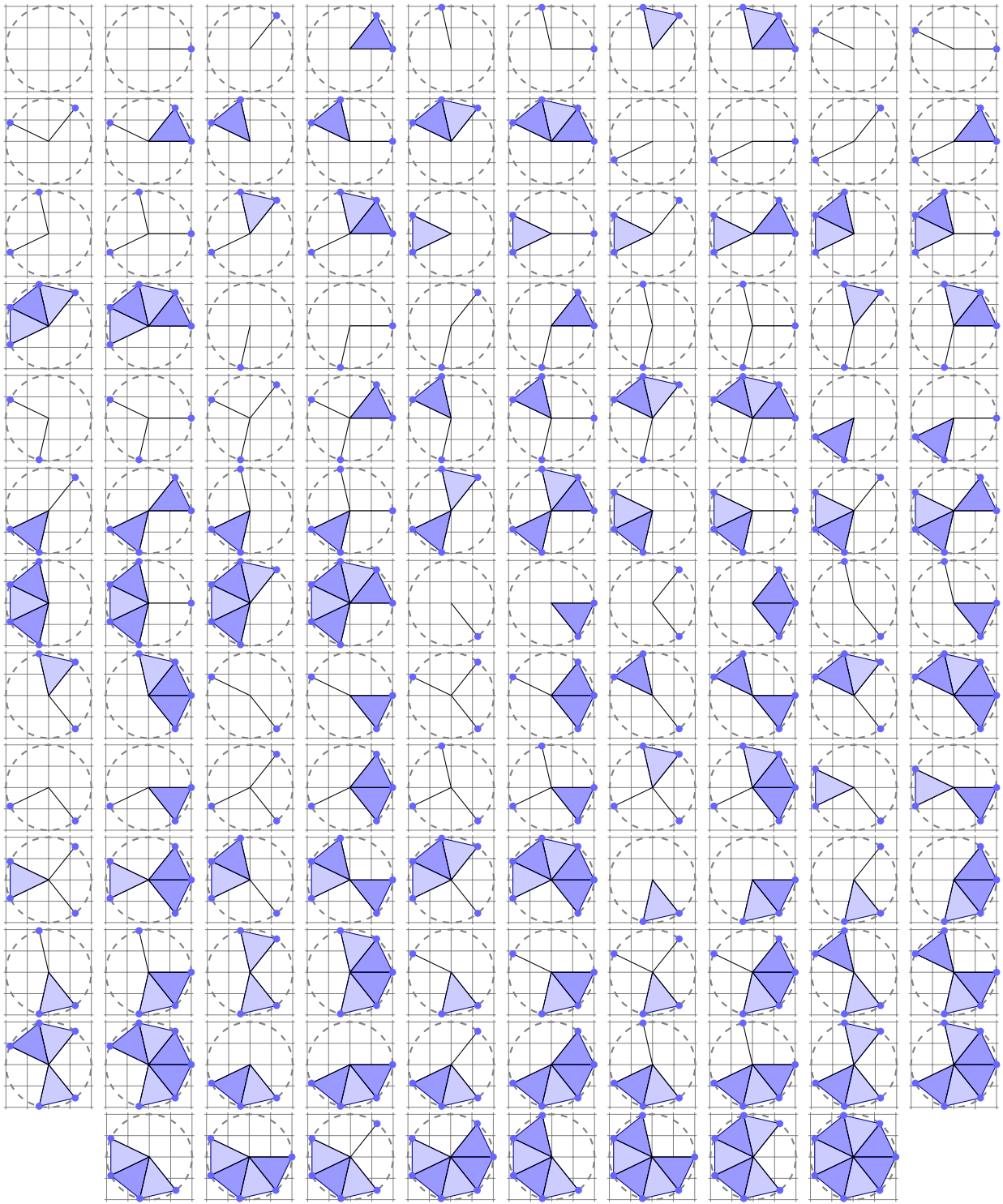

\centering
\input{./tikz/0-small.tikz}
\input{./tikz/1-small.tikz}
\input{./tikz/2-small.tikz}
\input{./tikz/3-small.tikz}
\input{./tikz/4-small.tikz}
\input{./tikz/5-small.tikz}
\input{./tikz/6-small.tikz}
\input{./tikz/7-small.tikz}
\input{./tikz/8-small.tikz}
\input{./tikz/9-small.tikz}
\input{./tikz/10-small.tikz}
\input{./tikz/11-small.tikz}
\input{./tikz/12-small.tikz}
\input{./tikz/13-small.tikz}
\input{./tikz/14-small.tikz}
\input{./tikz/15-small.tikz}
\input{./tikz/16-small.tikz}
\input{./tikz/17-small.tikz}
\input{./tikz/18-small.tikz}
\input{./tikz/19-small.tikz}
\input{./tikz/20-small.tikz}
\input{./tikz/21-small.tikz}
\input{./tikz/22-small.tikz}
\input{./tikz/23-small.tikz}
\input{./tikz/24-small.tikz}
\input{./tikz/25-small.tikz}
\input{./tikz/26-small.tikz}
\input{./tikz/27-small.tikz}
\input{./tikz/28-small.tikz}
\input{./tikz/29-small.tikz}
\input{./tikz/30-small.tikz}
\input{./tikz/31-small.tikz}
\input{./tikz/32-small.tikz}
\input{./tikz/33-small.tikz}
\input{./tikz/34-small.tikz}
\input{./tikz/35-small.tikz}
\input{./tikz/36-small.tikz}
\input{./tikz/37-small.tikz}
\input{./tikz/38-small.tikz}
\input{./tikz/39-small.tikz}
\input{./tikz/40-small.tikz}
\input{./tikz/41-small.tikz}
\input{./tikz/42-small.tikz}
\input{./tikz/43-small.tikz}
\input{./tikz/44-small.tikz}
\input{./tikz/45-small.tikz}
\input{./tikz/46-small.tikz}
\input{./tikz/47-small.tikz}
\input{./tikz/48-small.tikz}
\input{./tikz/49-small.tikz}
\input{./tikz/50-small.tikz}
\input{./tikz/51-small.tikz}
\input{./tikz/52-small.tikz}
\input{./tikz/53-small.tikz}
\input{./tikz/54-small.tikz}
\input{./tikz/55-small.tikz}
\input{./tikz/56-small.tikz}
\input{./tikz/57-small.tikz}
\input{./tikz/58-small.tikz}
\input{./tikz/59-small.tikz}
\input{./tikz/60-small.tikz}
\input{./tikz/61-small.tikz}
\input{./tikz/62-small.tikz}
\input{./tikz/63-small.tikz}
\input{./tikz/64-small.tikz}
\input{./tikz/65-small.tikz}
\input{./tikz/66-small.tikz}
\input{./tikz/67-small.tikz}
\input{./tikz/68-small.tikz}
\input{./tikz/69-small.tikz}
\input{./tikz/70-small.tikz}
\input{./tikz/71-small.tikz}
\input{./tikz/72-small.tikz}
\input{./tikz/73-small.tikz}
\input{./tikz/74-small.tikz}
\input{./tikz/75-small.tikz}
\input{./tikz/76-small.tikz}
\input{./tikz/77-small.tikz}
\input{./tikz/78-small.tikz}
\input{./tikz/79-small.tikz}
\input{./tikz/80-small.tikz}
\input{./tikz/81-small.tikz}
\input{./tikz/82-small.tikz}
\input{./tikz/83-small.tikz}
\input{./tikz/84-small.tikz}
\input{./tikz/85-small.tikz}
\input{./tikz/86-small.tikz}
\input{./tikz/87-small.tikz}
\input{./tikz/88-small.tikz}
\input{./tikz/89-small.tikz}
\input{./tikz/90-small.tikz}
\input{./tikz/91-small.tikz}
\input{./tikz/92-small.tikz}
\input{./tikz/93-small.tikz}
\input{./tikz/94-small.tikz}
\input{./tikz/95-small.tikz}
\input{./tikz/96-small.tikz}
\input{./tikz/97-small.tikz}
\input{./tikz/98-small.tikz}
\input{./tikz/99-small.tikz}
\input{./tikz/100-small.tikz}
\input{./tikz/101-small.tikz}
\input{./tikz/102-small.tikz}
\input{./tikz/103-small.tikz}
\input{./tikz/104-small.tikz}
\input{./tikz/105-small.tikz}
\input{./tikz/106-small.tikz}
\input{./tikz/107-small.tikz}
\input{./tikz/108-small.tikz}
\input{./tikz/109-small.tikz}
\input{./tikz/110-small.tikz}
\input{./tikz/111-small.tikz}
\input{./tikz/112-small.tikz}
\input{./tikz/113-small.tikz}
\input{./tikz/114-small.tikz}
\input{./tikz/115-small.tikz}
\input{./tikz/116-small.tikz}
\input{./tikz/117-small.tikz}
\input{./tikz/118-small.tikz}
\input{./tikz/119-small.tikz}
\input{./tikz/120-small.tikz}
\input{./tikz/121-small.tikz}
\input{./tikz/122-small.tikz}
\input{./tikz/123-small.tikz}
\input{./tikz/124-small.tikz}
\input{./tikz/125-small.tikz}
\input{./tikz/126-small.tikz}
\input{./tikz/127-small.tikz}

\caption{Geometric representation of all binary 7-tuple.}
\label{figure-binary-panel}
\end{figure*}

\begin{figure*}
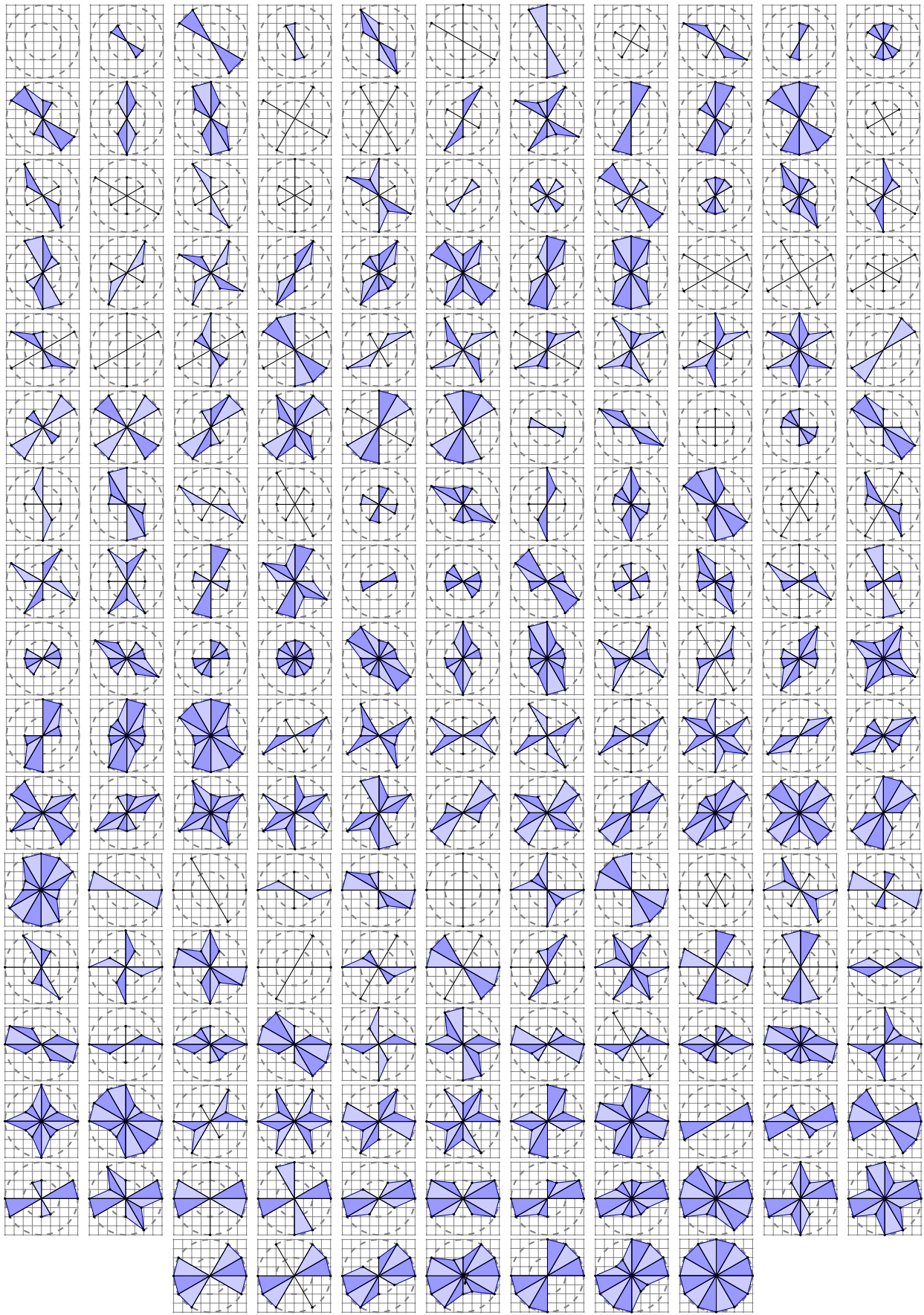

\centering

\input{./tikz/t-000000000000.tikz}
\input{./tikz/t-000011000011.tikz}
\input{./tikz/t-000022000022.tikz}
\input{./tikz/t-000110000110.tikz}
\input{./tikz/t-000121000121.tikz}
\input{./tikz/t-000202000202.tikz}
\input{./tikz/t-000220000220.tikz}
\input{./tikz/t-001001001001.tikz}
\input{./tikz/t-001012001012.tikz}
\input{./tikz/t-001100001100.tikz}
\input{./tikz/t-001111001111.tikz}
\input{./tikz/t-001122001122.tikz}
\input{./tikz/t-001210001210.tikz}
\input{./tikz/t-001221001221.tikz}
\input{./tikz/t-002002002002.tikz}
\input{./tikz/t-002020002020.tikz}
\input{./tikz/t-002101002101.tikz}
\input{./tikz/t-002112002112.tikz}
\input{./tikz/t-002200002200.tikz}
\input{./tikz/t-002211002211.tikz}
\input{./tikz/t-002222002222.tikz}
\input{./tikz/t-010010010010.tikz}
\input{./tikz/t-010021010021.tikz}
\input{./tikz/t-010102010102.tikz}
\input{./tikz/t-010120010120.tikz}
\input{./tikz/t-010201010201.tikz}
\input{./tikz/t-010212010212.tikz}
\input{./tikz/t-011000011000.tikz}
\input{./tikz/t-011011011011.tikz}
\input{./tikz/t-011022011022.tikz}
\input{./tikz/t-011110011110.tikz}
\input{./tikz/t-011121011121.tikz}
\input{./tikz/t-011202011202.tikz}
\input{./tikz/t-011220011220.tikz}
\input{./tikz/t-012001012001.tikz}
\input{./tikz/t-012012012012.tikz}
\input{./tikz/t-012100012100.tikz}
\input{./tikz/t-012111012111.tikz}
\input{./tikz/t-012122012122.tikz}
\input{./tikz/t-012210012210.tikz}
\input{./tikz/t-012221012221.tikz}
\input{./tikz/t-020002020002.tikz}
\input{./tikz/t-020020020020.tikz}
\input{./tikz/t-020101020101.tikz}
\input{./tikz/t-020112020112.tikz}
\input{./tikz/t-020200020200.tikz}
\input{./tikz/t-020211020211.tikz}
\input{./tikz/t-020222020222.tikz}
\input{./tikz/t-021010021010.tikz}
\input{./tikz/t-021021021021.tikz}
\input{./tikz/t-021102021102.tikz}
\input{./tikz/t-021120021120.tikz}
\input{./tikz/t-021201021201.tikz}
\input{./tikz/t-021212021212.tikz}
\input{./tikz/t-022000022000.tikz}
\input{./tikz/t-022011022011.tikz}
\input{./tikz/t-022022022022.tikz}
\input{./tikz/t-022110022110.tikz}
\input{./tikz/t-022121022121.tikz}
\input{./tikz/t-022202022202.tikz}
\input{./tikz/t-022220022220.tikz}
\input{./tikz/t-100001100001.tikz}
\input{./tikz/t-100012100012.tikz}
\input{./tikz/t-100100100100.tikz}
\input{./tikz/t-100111100111.tikz}
\input{./tikz/t-100122100122.tikz}
\input{./tikz/t-100210100210.tikz}
\input{./tikz/t-100221100221.tikz}
\input{./tikz/t-101002101002.tikz}
\input{./tikz/t-101020101020.tikz}
\input{./tikz/t-101101101101.tikz}
\input{./tikz/t-101112101112.tikz}
\input{./tikz/t-101200101200.tikz}
\input{./tikz/t-101211101211.tikz}
\input{./tikz/t-101222101222.tikz}
\input{./tikz/t-102010102010.tikz}
\input{./tikz/t-102021102021.tikz}
\input{./tikz/t-102102102102.tikz}
\input{./tikz/t-102120102120.tikz}
\input{./tikz/t-102201102201.tikz}
\input{./tikz/t-102212102212.tikz}
\input{./tikz/t-110000110000.tikz}
\input{./tikz/t-110011110011.tikz}
\input{./tikz/t-110022110022.tikz}
\input{./tikz/t-110110110110.tikz}
\input{./tikz/t-110121110121.tikz}
\input{./tikz/t-110202110202.tikz}
\input{./tikz/t-110220110220.tikz}
\input{./tikz/t-111001111001.tikz}
\input{./tikz/t-111012111012.tikz}
\input{./tikz/t-111100111100.tikz}
\input{./tikz/t-111111111111.tikz}
\input{./tikz/t-111122111122.tikz}
\input{./tikz/t-111210111210.tikz}
\input{./tikz/t-111221111221.tikz}
\input{./tikz/t-112002112002.tikz}
\input{./tikz/t-112020112020.tikz}
\input{./tikz/t-112101112101.tikz}
\input{./tikz/t-112112112112.tikz}
\input{./tikz/t-112200112200.tikz}
\input{./tikz/t-112211112211.tikz}
\input{./tikz/t-112222112222.tikz}
\input{./tikz/t-120010120010.tikz}
\input{./tikz/t-120021120021.tikz}
\input{./tikz/t-120102120102.tikz}
\input{./tikz/t-120120120120.tikz}
\input{./tikz/t-120201120201.tikz}
\input{./tikz/t-120212120212.tikz}
\input{./tikz/t-121000121000.tikz}
\input{./tikz/t-121011121011.tikz}
\input{./tikz/t-121022121022.tikz}
\input{./tikz/t-121110121110.tikz}
\input{./tikz/t-121121121121.tikz}
\input{./tikz/t-121202121202.tikz}
\input{./tikz/t-121220121220.tikz}
\input{./tikz/t-122001122001.tikz}
\input{./tikz/t-122012122012.tikz}
\input{./tikz/t-122100122100.tikz}
\input{./tikz/t-122111122111.tikz}
\input{./tikz/t-122122122122.tikz}
\input{./tikz/t-122210122210.tikz}
\input{./tikz/t-122221122221.tikz}
\input{./tikz/t-200002200002.tikz}
\input{./tikz/t-200020200020.tikz}
\input{./tikz/t-200101200101.tikz}
\input{./tikz/t-200112200112.tikz}
\input{./tikz/t-200200200200.tikz}
\input{./tikz/t-200211200211.tikz}
\input{./tikz/t-200222200222.tikz}
\input{./tikz/t-201010201010.tikz}
\input{./tikz/t-201021201021.tikz}
\input{./tikz/t-201102201102.tikz}
\input{./tikz/t-201120201120.tikz}
\input{./tikz/t-201201201201.tikz}
\input{./tikz/t-201212201212.tikz}
\input{./tikz/t-202000202000.tikz}
\input{./tikz/t-202011202011.tikz}
\input{./tikz/t-202022202022.tikz}
\input{./tikz/t-202110202110.tikz}
\input{./tikz/t-202121202121.tikz}
\input{./tikz/t-202202202202.tikz}
\input{./tikz/t-202220202220.tikz}
\input{./tikz/t-210001210001.tikz}
\input{./tikz/t-210012210012.tikz}
\input{./tikz/t-210100210100.tikz}
\input{./tikz/t-210111210111.tikz}
\input{./tikz/t-210122210122.tikz}
\input{./tikz/t-210210210210.tikz}
\input{./tikz/t-210221210221.tikz}
\input{./tikz/t-211002211002.tikz}
\input{./tikz/t-211020211020.tikz}
\input{./tikz/t-211101211101.tikz}
\input{./tikz/t-211112211112.tikz}
\input{./tikz/t-211200211200.tikz}
\input{./tikz/t-211211211211.tikz}
\input{./tikz/t-211222211222.tikz}
\input{./tikz/t-212010212010.tikz}
\input{./tikz/t-212021212021.tikz}
\input{./tikz/t-212102212102.tikz}
\input{./tikz/t-212120212120.tikz}
\input{./tikz/t-212201212201.tikz}
\input{./tikz/t-212212212212.tikz}
\input{./tikz/t-220000220000.tikz}
\input{./tikz/t-220011220011.tikz}
\input{./tikz/t-220022220022.tikz}
\input{./tikz/t-220110220110.tikz}
\input{./tikz/t-220121220121.tikz}
\input{./tikz/t-220202220202.tikz}
\input{./tikz/t-220220220220.tikz}
\input{./tikz/t-221001221001.tikz}
\input{./tikz/t-221012221012.tikz}
\input{./tikz/t-221100221100.tikz}
\input{./tikz/t-221111221111.tikz}
\input{./tikz/t-221122221122.tikz}
\input{./tikz/t-221210221210.tikz}
\input{./tikz/t-221221221221.tikz}
\input{./tikz/t-222002222002.tikz}
\input{./tikz/t-222020222020.tikz}
\input{./tikz/t-222101222101.tikz}
\input{./tikz/t-222112222112.tikz}
\input{./tikz/t-222200222200.tikz}
\input{./tikz/t-222211222211.tikz}
\input{./tikz/t-222222222222.tikz}

\caption{Geometric representation of selected  words of the discussed ternary Golay transform.}
\label{figure-selected-ternary-flowers}
\end{figure*}

\section*{Acknowledgements}
The second author acknowledges the support
from CNPq.

\onecolumn

{\small
\singlespacing
\bibliographystyle{siam}
\bibliography{geom.bib}
}

\end{document}